\documentclass{article}
\usepackage{graphicx}
\usepackage{epstopdf}
\usepackage{lmodern}
\title{\textbf{EFFECTIVE POTENTIAL AND GEODESIC MOTION IN KERR-de SITTER SPACE-TIME}}
\author{P.C. POUDEL$^{*}$ AND U. KHANAL$^{*}$\\${*}$ Central Department Of Physics, T.U., Kirtipur}
\date{\today}

\chardef\bslchar=`\\ 

\providecommand{\qedsymbol}{\leavevmode
  \hbox to.77778em{%
  \hfil\vrule
  \vbox to.675em{\hrule width.6em\vfil\hrule}%
  \vrule\hfil}}

\def\latex/{{\protect\LaTeX}}

\begin{document}
\maketitle
\textbf{Abstract:}{{\quad In the present work, geodesic trajectories in Kerr-de Sitter geometry is analyzed. From the
   mathematical solution of Lagrangian formalism appropriate to motions in the equatorial plane
  (for which $\dot{\theta}=0$ and $\theta=$ (constant)$=\pi/2$) can give potential energy of massive
  and massless particles for rotating axisymetric black hole. From this, for a particular value of  cosmological constant, Kerr parameter, mass, angular momentum and impact parameter; variation of potential with distance can be found.
  Similarly, for a particular value of cosmological constant, mass and Kerr parameter; variation of velocity with
  distance can be found.}}\\\\
  \qquad \textbf{Keywords}: \,Effective Potential, Geodesic Motion, Kerr-de Sitter Space-Time, Hamiltonian Formalism, Cosmological Constant, Kerr Parameter, Mass of Galaxies

\markboth{Effective Potential And Geodesic Motion In Kerr-de Sitter Space-Time:\,\,\textsf{P.C. Poudel \& U.Khanal}}{Effective Potential And Geodesic Motion In Kerr-de Sitter Space-Time:\,\,\textsf{P.C. Poudel \& U.Khanal}}


\section{Introduction}
Cosmological constant $\Lambda$ was introduced by Einstein to balance the evolutionary models with repulsion to set a steady state. He abandoned it later as a blunder. When Hubble expansion was seen, it has now returned with a vengeance as the accelerating expansion to contribute 75\% of the density of the universe as dark energy. $\Lambda$ can also be introduced as the vacuum energy that is required to drive inflation (Akcay, 2009). The accelerating expansion may even be interpreted as the continuation of inflation, possibly at a slower rate than in the early universe.

The concept of space and time can be made from the study of black holes of the nature. A black hole is supposed to possess three physical properties: mass, angular momentum and charge. Charged black hole is expected to absorb the opposite charge and become neutral. The trajectory, of massive and massless particles in various geometries, is described by the geodesics. In particular, the behaviour of massive and massless objects around a spherically symmetric gravitating body is described by Schwarzschild formalism. Schwarzschild space-time is a solution obtained from the Einstein's field equations that is static. When we introduce $\Lambda$ in Schwarzschild space-time solution, we obtain Schwarzschild-de Sitter space-time. Unlike Schwarzschild solution, Kerr solution is an axi-symmetric solution of Einstein's field equations corresponding to a rotating black hole. A rotating black hole in asymptotically de-Sitter space-time can be described by Kerr-de Sitter geometry.

\section{The Geodesics in the Equatorial Plane in Kerr-de Sitter space-time}

The geodesics in
the equatorial plane can be delineated in very much as in
schwarzschild: the energy and angular momentum integrals will
suffice to reduce  the problem to one of the quadratures. But two
essentials differences must be kept in mind. First, a distinction
should be made between direct and retrograde orbits whose rotation
about the axis of symmetry are in the same sense or in opposite
sense to that of the black hole. And second, the co-ordinate
$\phi$, like the co-ordinate t, is not a good co-ordinate for
describing what really happens with respect to a co-moving
observer: a trajectory approaching the horizon ( at $r_{+}$ or
$r_{-}$ ) will spiral round the black hole an infinite co-ordinate
time t to cross the horizon; and neither will be the experience of
the co-moving observer.
\\
The lagrangian appropriate to motions in the equatorial plane (
for which $\dot{\theta}=0$ and $\theta$= a
constant=$\frac{\pi}{2}$ ) is
\begin{displaymath}
\mathcal{L}=\frac{1}{2}\,g_{\mu\nu}\,\frac{dx^{\mu}}{d\tau}\,\frac{dx^{\nu}}{d\tau}
\end{displaymath}
\begin{eqnarray}
2\,\mathcal{L}=\frac{\Delta_r-\Delta_\theta a^2\,sin^2
\theta}{\rho^2 \Xi^2}\dot{t}^2 + \frac{2a\,sin^2 \theta}{\rho^2
\Xi^2}\Big[\big(r^2+a^2\big)\Delta_\theta-\Delta_r\Big]\dot{t}\dot{\phi}-\frac{\rho^2}{\Delta_r}\dot{r}^2-\frac{\rho^2}{\Delta_\theta}\dot{\theta}^2 \nonumber \\
-\frac{sin^2\theta}{\rho^2\Xi^2}\Big[\big(r^2+a^2\big)^2\Delta_\theta-\Delta_r
a^2 sin^2\theta\Big]\dot{\phi}^2\end{eqnarray}
\\
and we deduce from it that the generalized momenta are given by,
\begin{equation}\label{4.1}
p_{t}=\frac{\partial \mathcal{L}}{\partial \dot{t}} =
\frac{\Delta_r-\Delta_\theta a^2\,sin^2 \theta}{\rho^2
\Xi^2}\dot{t} + \frac{a\,sin^2 \theta}{\rho^2
\Xi^2}\Big[\big(r^2+a^2\big)\Delta_\theta-\Delta_r\Big]\dot{\phi}
=E=constant
\end{equation}
\begin{eqnarray}\label{4.2}
-p_{\phi}=-\frac{\partial \mathcal{L}}{\partial\dot{\phi}}
=\frac{sin^2\theta}{\rho^2\Xi^2}\Big[\big(r^2+a^2\big)^2\Delta_\theta-\Delta_r
a^2 sin^2\theta\Big]\dot{\phi}-\frac{a\,sin^2 \theta}{\rho^2
\Xi^2}\Big[\big(r^2+a^2\big)\Delta_\theta-\Delta_r\Big]\dot{t}
\nonumber \\
= L = constant
\end{eqnarray}
\\
\begin{equation}
-p_{r}=-\frac{\partial \mathcal{L}}{\partial
\dot{r}}=\frac{\rho^2}{\Delta_{r}}\dot{r}
\end{equation}
\\ and
\begin{equation}
-p_{\theta}=-\frac{\partial \mathcal{L}}{\partial
\dot{\theta}}=\frac{\rho^2}{\Delta_{\theta}}\dot{\theta}
\end{equation}\\
where we have used superior dots to denote the differentiation
with respect to an affine parameter $\tau$.
\\\\
The constancy of $p_{t}$ and $p_{\phi}$ follows from the
independence of the lagrangian on $t$ and $\phi$ which, in turn,
is a manifestation of the stationary and the axisymmetric
character of Kerr-de sitter geometry.
\\\\
The Hamiltonian is given by
\begin{displaymath}
H=p_{t}\,\dot{t}-
\big(p_{\phi}\,\dot{\phi}+p_{r}\,\dot{r}+p_{\theta}
\dot{\theta}\big)-\mathcal{L}
\end{displaymath}
and from the independence of the Hamiltonian on t, we deduce that
\begin{eqnarray}
2\,H=\bigg\{\frac{\Delta_r-\Delta_\theta a^2\,sin^2 \theta}{\rho^2
\Xi^2}\dot{t} + \frac{a\,sin^2 \theta}{\rho^2
\Xi^2}\Big[\big(r^2+a^2\big)\Delta_\theta-\Delta_r\Big]\dot{\phi}\bigg\}\dot{t}-\frac{sin^2\theta}{\rho^2
\Xi^2}\bigg\{\Big[\big(r^2+a^2\big)^2\Delta_\theta \nonumber \\
-\Delta_r a^2
sin^2\theta\Big]\dot{\phi}-a\Big[\big(r^2+a^2\big)\Delta_\theta-\Delta_r\Big]\dot{t}\bigg\}\dot{\phi}-\frac{\rho^2}{\Delta_{r}}\dot{r}^2
-\frac{\rho^2}{\Delta_{\theta}}\dot{\theta}^2 \nonumber
\end{eqnarray}
\begin{equation}\label{4.3}
2\,H=E\,\dot{t}-L\,\dot{\phi}-\frac{\rho^{2}}{\Delta_r}\,\dot{r}^{2}-\frac{\rho^2}{\Delta_\theta}\dot{\theta}^2=\delta_1=constant
\end{equation}
we can set, for time like geodesics,
\begin{displaymath}
\delta_{1}=1
\end{displaymath}
and for null geodesics,
\begin{displaymath}
\delta_{1}=0
\end{displaymath}
(Setting $\delta_1=1$ for time like geodesics requires $E$ to be
interpreted as the specific energy or energy per unit mass)\\
solving equations (\ref{4.1}) and (\ref{4.2}), for $\dot{\phi}$
and $\dot{t}$, we obtain
\begin{equation}
\dot{\phi}=\frac{\Xi^2}{\rho^2 \Delta_\theta \Delta_r
sin^2\theta}\,\big[a\Delta_\theta\{E(r^2+a^2)-a
L\}-\frac{\Delta_r}{sin^2\theta}(aE sin^2\theta- L)\big]
\end{equation}
\begin{equation}
\dot{t}=\frac{\Xi^2}{\rho^2 \Delta_\theta \Delta_r
sin^2\theta}\,\big[(r^2+a^2)\Delta_\theta\{E(r^2+a^2)-aL\}-a\Delta_r(aE
sin^2\theta- L)\big]
\end{equation}
and inserting these solutions in equation (\ref{4.3})
\begin{eqnarray}\label{4.4}
\dot{r}^{2}= \frac{\Xi^2}{\rho^4
\Delta_\theta}\Big[E^2\big\{(r^2+a^2)^2\Delta_\theta-\Delta_r
a^2sin^2\theta
\big\}-\frac{L^2}{sin^2\theta}(\Delta_r-a^2\Delta_\theta
sin^2\theta)\nonumber \\
-2aEL\big\{(r^2+a^2)\Delta_\theta
-\Delta_r\big\}\Big]-\frac{\Delta_r}{\Delta_\theta}\dot{\theta}^2-\frac{\Delta_r}{\rho^2}\delta_1
\end{eqnarray}
\\
Now, to find velocity of test particle, we know \\
\begin{equation} \Omega=\frac{d\phi}{dt}=\frac
{\dot{\phi}}{\dot{t}}\end{equation}\\
and \begin{equation}
\omega=\frac{a}{\Sigma^2}\big[(r^2+a^2)\Delta_\theta-\Delta_r\big]\end{equation}\\
where $\Sigma^2=(r^2+a^2)^2\Delta_\theta-\Delta_r a^2
sin^2\theta$\\\\
The rotational velocity of test particle in the orbit around the
central mass is \begin{equation} V_\varphi =\frac{\Sigma^2
sin^2\theta}{\rho^2\sqrt{\Delta_r
\Delta_\theta}}(\Omega-\omega)\end{equation} The quantitative
feature of geodesic motion in the equatorial plane is very
illustrative, henceforth $\dot{\theta}=0$ and $\theta=\pi/2$ and
equation (\ref{4.4}) simplifies to \begin{equation}\label{4.5}
\dot{r}^2=\frac
{\Xi^2}{r^4}\Big[\big\{E(r^2+a^2)-aL\big\}^2-\Delta_r(aE-L)^2\Big]-
\frac{\Delta_r}{r^2}\delta_1 \end{equation} \\
Or, \begin{equation} \dot{r}^2=\frac
{1}{r^4}\Big[\Xi^2\big\{E(r^2+a^2)-aL\big\}^2-\Delta_r(r^2\delta_1+K)\Big]
\end{equation}\\
where, $K=\Xi^2(aE-L)^2= constant$
\\

\subsection{The Null Geodesics}
As we have noted, $\delta_1=0$ for null geodesics and the radial
equation (\ref{4.5}) becomes \begin{equation} \dot{r}^2 =
E^2+\frac{\Lambda}{3}(L-aE)^2-\frac{1}{r^2}(L-aE)\big\{L+aE-\frac{\Lambda}{3}a^2(L-aE)\big\}+\frac{2M}{r^3}(L-aE)^2+\frac{1}{r^4}(.....)
\nonumber \end{equation}\\ i.e.  \begin{equation}\label{4.6}
\dot{r}^2 =
E^2+\frac{\Lambda}{3}(L-aE)^2-\frac{1}{r^2}(L-aE)\big\{L+aE-\frac{\Lambda}{3}a^2(L-aE)\big\}+\frac{2M}{r^3}(L-aE)^2
 \end{equation} \\
 In our further considerations, it will be more convenient to
 distinguish the geodesics by the impact parameter $D=\frac{L}{E}$
 rather than by $L$.\\
\\
\textbf{The special case: when $D=a$} \\
We observe that geodesics with impact parameter $D=a$ and $L=aE$
play, in the present context, the same as the geodesics in the
Schwarzschild and in the Reissner-Nordstorm geometry. Thus in the
case, equation (\ref{4.6}) reduce to \begin{equation} \dot{r}=\pm
E, ~~ \dot{t}=\frac{\Xi^2(r^2+a^2)}{\Delta_r}E
~~and~~\dot{\phi}=\frac{\Xi^2 a}{\Delta_r}E \nonumber
\end{equation}\\
The radial coordinate is described uniformly with respect to the
affine parameter while the equation governing t and $\phi$ are
\begin{displaymath}
\frac{dt}{dr}=\pm\frac{\Xi^2(r^2+a^2)}{\Delta_r}~~~
and~~~~\frac{d\phi}{dt}=\pm \frac{a}{r^2+a^2}\end{displaymath}\\
In general it is clear that we must distinguish orbits with impact
parameters greater than or less than a certain critical value
$D_c$, which will in turn be different for the direct and
retrograde orbits. For $D=D_c$, the geodesic equations allow an
unstable circular orbit of radius $r_c$ (say). For $D>D_c$ we
shall have orbits of two kinds: those of the first kind which
arriving from infinity, have perihelion distances greater than
$r_c$; and those of the second kind which having apehelion
distances less than $r_c$, terminate at the singularity $r=0$. For
$D=D_c$ the orbits of two kinds coalesce: they both spiral,
indefinitely about some unstable circular orbit at $r=r_c$. For
$D<D_c$, there are orbits of one kind: arriving from infinity,
they cross both
horizons and terminate at the singularity.\\\\
The equations determining the radius $r_c$ of the unstable
circular `photon orbits' are
\begin{equation}\label{4.7} E^2+\frac{\Lambda}{3}(L-aE)^2-\frac{1}{r^2}(L-aE)\big\{L+aE-\frac{\Lambda a^2}{3}(L-aE)\big\}+\frac{2M}{r^3}(L-aE)^2=0
 \end{equation}\\
and \begin{equation} \frac{2}{r^3}(L-aE)\big\{L+aE-\frac{\Lambda
a^2}{3}(L-aE)\big\}-\frac{6M}{r^4}(L-aE)^2=0
 \end{equation} \\
 From these equations, we conclude that \begin{equation}
 r_c=\frac{3M(L-aE)}{L+aE-\frac{\Lambda}{3}a^2(L-aE)}=\frac{3M(D_c-a)}{D_c+a-\frac{\Lambda}{3}a^2(D_c-a)}\end{equation}
 Inserting the last relation in the equation (\ref{4.7}), we find
 that the equation can be reduced to \begin{eqnarray}
 D_c^3\Big[\big(1-\frac{\Lambda
 a^2}{3}\big)^3-9M^2\Lambda\Big]+3D_c^2a\Big[\big(1-\frac{\Lambda a^2}{3}\big)^2\big(1+\frac{\Lambda
 a^2}{3}\big)+9M^2\Lambda\Big] \nonumber \\
 +3D_c\Big[a^2\big(1-\frac{\Lambda a^2}{3}\big)\big(1+\frac{\Lambda
 a^2}{3}\big)^2-9M^2(1+\Lambda a^2)\Big] \nonumber \\
 +27M^2a\big(1+\frac{\Lambda a^2}{3}\big)+a^3\big(1+\frac{\Lambda
 a^2}{3}\big)^3=0 \end{eqnarray} \\
This cubic equation can be solved by the standard method by
changing variable \begin{displaymath} D_c = y -
\frac{a\Big[\big(1-\frac{\Lambda
a^2}{3}\big)^2\big(1+\frac{\Lambda
 a^2}{3}\big)+9M^2\Lambda\Big]}{\Big[\big(1-\frac{\Lambda
 a^2}{3}\big)^3-9M^2\Lambda\Big]} \end{displaymath}
 to obtain the cubic equation \begin{displaymath}
 (D_c+z_1)^3=27 M^2A(D_c-z_2)\end{displaymath}
 i.e.,\begin{equation}\label{4.8} y^3-27M^2 A y+54M^2 A B=0 \end{equation}
 where, for simplicity if we consider \begin{displaymath}
 p=\big(1-\frac{\Lambda a^2}{3}\big),q=\big(1+\frac{\Lambda
 a^2}{3}\big) \nonumber \end{displaymath} \\
Then, \begin{displaymath}z_1=\frac{a\Big[\big(1-\frac{\Lambda
a^2}{3}\big)^2\big(1+\frac{\Lambda
 a^2}{3}\big)+9M^2\Lambda\Big]}{\Big[\big(1-\frac{\Lambda
 a^2}{3}\big)^3-9M^2\Lambda\Big]}=\frac{a(p^2q)+9M^2\Lambda}{p^3-9M^2\Lambda},\nonumber \end{displaymath} \\
\begin{displaymath}z_2=\frac{a}{M^2}\frac{\big\{(a^2q^3+27M^2q)(p^3-9M^2\Lambda)^2-a^2(p^2q+9M^2\Lambda)^2\big\}}{\big\{p^3+4\Lambda
a^2p-9M^2\Lambda\big\}},\nonumber \end{displaymath}\\
\begin{displaymath}A=\frac{\big\{p^3+4\Lambda
a^2p-9M^2\Lambda\big\}}{(p^3-9M^2\Lambda)^2} \nonumber \end{displaymath}\\
and,\begin{displaymath} B=\frac{z_1+z_2}{2}\end{displaymath}\\
We must now distinguish the $a>0$ and $a<0$ corresponding to the
direct and retrograde orbits. For $a>0$ \begin{equation}
y_+=-\frac{6M\Big[\big(1-\frac{\Lambda
a^2}{3}\big)^3+4a^2\Lambda\big(1-\frac{\Lambda
a^2}{3}\big)\Big]^\frac{1}{2}}{\big(1-\frac{\Lambda
a^2}{3}\big)^3-9M^2\Lambda}cos\big(\varphi +2\pi/3\big)
\end{equation}\\
and for $-a=|a|>0$ \begin{equation}
y_-=-\frac{6M\Big[\big(1-\frac{\Lambda
a^2}{3}\big)^3+4a^2\Lambda\big(1-\frac{\Lambda
a^2}{3}\big)\Big]^\frac{1}{2}}{\big(1-\frac{\Lambda
a^2}{3}\big)^3-9M^2\Lambda}cos\varphi
\end{equation}\\
where,\begin{displaymath}cos3\varphi=\frac{|a|}{M}\frac{(z_1+z_2)}{2aA^{1/2}}
=\frac{|a|}{M}\Bigg[\frac{\big(1-\frac{\Lambda
a^2}{3}\big)\big(1+\frac{\Lambda a^2}{3}\big)-9M^2\Lambda
\bigg\{\big(1-\frac{\Lambda
a^2}{3}\big)^2-\frac{4a^2\Lambda}{3}\bigg\}}{\big(1-\frac{\Lambda
a^2}{3}\big)^3-9M^2\Lambda+4a^2\Lambda\big(1-\frac{\Lambda
a^2}{3}\big)}\Bigg]\end{displaymath} and, for $a=0$ we have \quad
$\varphi=\frac{5\pi}{6}$ so that
\begin{equation} D_c=\frac{3\sqrt{3}M}{\sqrt{1-9M^2\Lambda}}, \quad
r_c=3M \end{equation}\\
Turning to the equations governing to the orbits when the impact
parameter has the critical value $D_c$, and the expression on the
left hand side of equation (\ref{4.7}) allows a double root, we
find that the equation can be reduced to the form
\begin{equation}
\dot{r}^2=\Big(-\frac{\dot{u}}{u^2}\Big)^2=ME^2(D_c-a)^2(u-u_c)^2(2u+u_c)\end{equation} \nonumber \\
where,\begin{displaymath}u=\frac{1}{r} and \quad
u_c=\frac{1}{r_c}=\frac{L+aE-\frac{\Lambda
a^2}{3}(L-aE)}{3M(L-aE)}\end{displaymath} This gives
\begin{equation}\label{4.9}\dot{u}^2=ME^2u^4(D_c-a)^2(u-u_c)^2(2u+u_c)\end{equation}\\
Equation (\ref{4.9}) can be integrated directly to give
\begin{eqnarray}\label{4.10} \Big[E(D_c-a)\sqrt{M}\Big]\tau=\pm\int\frac{du}{u^2(u-u_c)(2u+u_c)^{1/2}}
\nonumber \\
=\pm\frac{1}{u_c^2}\bigg\{\frac{1}{u_c}(2u+u_c)^{1/2}+\frac{1}{\sqrt{3u_c}}lg\bigg|\frac{\sqrt{2u+u_c}-\sqrt{3u_c}}{\sqrt{2u+u_c}+\sqrt{3u_c}}\bigg|\bigg\}\end{eqnarray}
But if we wish to exhibit the orbit in the equatorial plane, we
may combine it with the equation \begin{equation}\label{4.11}
\dot{\phi}=\frac{Eu^2(3+a^2\Lambda)\big\{6Mu^3(a-D)+u^2(a^3\Lambda-\Lambda
a^2D+3D)+\Lambda(a-D)\big\}}{3(3u^2+3a^2u^4-6Mu^3-\Lambda-\Lambda
a^2u^2)} \end{equation}
 and from equations (\ref{4.9}) and (\ref{4.11}), we obtain
\begin{equation}\label{4.12}
\pm\frac{d\phi}{du}=\frac{\dot{\phi}}{\dot{u}}=\frac{(3+a^2\Lambda)\big\{6Mu^3(a-D)+u^2(a^3\Lambda-\Lambda
a^2D+3D)+\Lambda(a-D)\big\}}{3\sqrt{M}(D_c-a)(3u^2+3a^2u^4-6Mu^3-\Lambda-\Lambda
a^2u^2)(u-u_c)(2u+u_c)^{1/2}} \end{equation}\\
From equations (\ref{4.10}) and (\ref{4.12}) orbits are derived.
They exhibit the features we have already described. The nature of
orbits in general, can be visualized from the orbits with the
critical impact parameters illustrated.

\subsection{\textbf{Time Like Geodesics}}
For time like geodesics, equations for for $\dot{\phi}$ and
$\dot{t}$ remain unchanged, but equation (\ref{4.5}) with
$\delta_{1}=1$ is given by
\begin{eqnarray}\label{4.13}
\dot{r}^2=\quad E^2+\frac{\Lambda}{3}(L-aE)^2-\frac{1}{r^2}(L-aE)\big\{L+aE-\frac{\Lambda}{3}a^2(L-aE)\big\}
+\frac{2M}{r^3}(L-aE)^2-\frac{\Delta_r}{r^2}\end{eqnarray}
 where E is now to be
interpreted as the energy per unit mass of the particle describing
the trajectory.\\
\\
\textbf{i) The spacial case, $L$=$a$$E$},
\\ \\
Time like geodesics with $L$=$a$$E$, are of interest in that their
behavior as they cross the horizon is characteristic of the orbits
in general.\\
\\
When $L$=$a$$E$, equation (\ref{4.13}) becomes
\begin{equation}\label{4.14}
r^{2}\,\dot{r}^{2}=r^{2}E^2-\Delta_r
\end{equation}
while,
\begin{equation}
\dot{\phi}=\frac{\Xi^2aE}{\Delta_r}  \quad and \quad
\dot{t}=\frac{\Xi^2(r^2+a^2)}{\Delta_r}E
\end{equation}
equation (\ref{4.14}) on integration gives, for $E^{2}>1$,
\begin{equation}
\tau=\int\big[E^2-\frac{\Delta_r}{r^2}\big]^{-1/2}dr
\end{equation}

\textbf{ii) The circular and associated orbits:}
\\ \\
We now turn to a consideration of the radial equation (\ref{4.13})
in general. With the reciprocal radius $u(=\frac{1}{r})$ as the
independent variable, the equation takes the form
\begin{eqnarray}\label{4.15}
\dot{r}^2=\frac{-\dot{u}^{2}}{u^{4}}
=-1+E^2+\frac{a^2\Lambda}{3}+(aE-L)^2\frac{\Lambda}{3} +\frac{\Lambda}{3u^2}+2Mu+\big\{a^2E^2-L^2-a^2 \nonumber\\ +\frac{a^2\Lambda}{3}(aE-L)^2\big\}u^2+2M(aE-L)^2u^3\quad
\end{eqnarray}
\subsubsection{\textbf{Effective potential approprite for time like trajectories}}
In equation (\ref{4.13}) $\dot{r}^2$ is interpreted as kinetic energy. As we know
total energy is the sum of kinetic energy and potential energy, in equation (\ref{4.13})
potential energy can be given as
\begin{eqnarray}
V=\frac{\Delta_r}{r^2}-\frac{\Lambda}{3}(L-aE)^2+\frac{1}{r^2}(L-aE)\big\{L+aE-\frac{\Lambda}{3}a^2(L-aE)\big\}
-\frac{2M}{r^3}(L-aE)^2\end{eqnarray}
As in our considerations, it will be more convenient to find effective potentials
by the impact parameter
\begin{displaymath}
D=\frac{L}{E}
\end{displaymath}
Introducing impact parameter in above equation, effective potential in time like geodesic
becomes
\begin{eqnarray}
V=\frac{\Delta_r}{r^2}-\frac{\Lambda}{3}L^2(1-a/D)^2+\frac{1}{r^2}L^2\big\{(1-a^2/D^2)-\frac{\Lambda}{3}a^2(1-a/D)\big\}
-\frac{2M}{r^3}L^2(1-a/D)^2\end{eqnarray}
For some typical values of the parameters $\Lambda$, $D$, $L$ and $a$ we can display
potential-energy curve with the variation of distance from the center.
The minima in the potentials corresponds to the stable circular orbits while
maxima corresponds to unstable circular orbits. At the point of inflection the
last stable circular orbit occurs.\\ \\

\begin{figure}[h]
\vspace{0.0cm} \centering
\includegraphics[height=8.0cm, width=11cm]{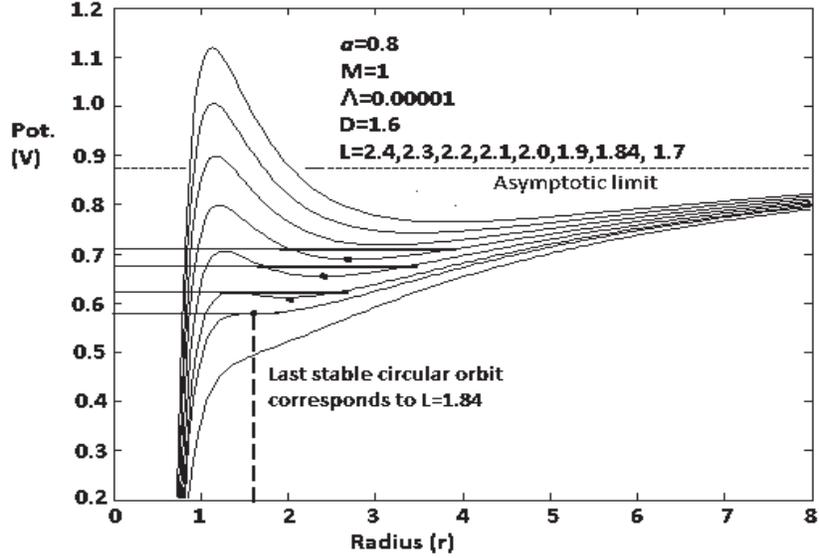}
\caption{Potential-energy curves appropriate for time-like trajectories in Kerr-de Sitter space time. }
\end{figure}

\subsubsection{\textbf{Rotational velocity in time like trajectory} }

As in the Schwarzschild and the Reissner-Nordstrom geometric, the
circular orbits play an important role in the classification of
the orbits. Besides, they are useful in providing simple examples
of orbits which exhibit the essential features at the same time;
and this is, after all, the reason for studying the geodesics.
\\\\
We seek then the values of $L$ and $E$ which a circular orbit at
some assigned radius, $r=\frac{1}{u}$, will have. When $L$ and $E$
have these values, the cubic polynomial on the right-hand side of
equation (\ref{4.15}) will have a double root.
\\
The conditions for the occurrence of a double root, after
substituting $x=L-aE$ are,
\begin{equation}\label{4.16}
\dot{r}^2r^4=f\,(r)=E^2r^4-r^2\Delta_r-2aEr^2x+(a^2-\Delta_r)x^2=0
\end{equation}
and differentiating equation (\ref{4.16}) with respect to r
\begin{equation}\label{4.17}
f'\,(r)=4E^2r^3-2r\Delta_r-r^2{\Delta_r}'-4aErx-{\Delta_r}'x^2=0
\end{equation}

Equations (\ref{4.16}) and (\ref{4.17}) can be written to give
\begin{equation}\label{4.21}
E^{2}r^4-\frac{r^3}{2}{\Delta_r}'+(\Delta_r-a^2-\frac{r}{2}{\Delta_r}')x^2=0
\end{equation}
\begin{equation}
E=\frac{1}{2ar^2x}\Big[(2a^{2}-2\Delta_r+\frac{r{\Delta_r}'}{2})x^2+\frac{r^3}{2}{\Delta_r}'-r^2\Delta_r\Big]
\end{equation}
after substituting this value of $E$ in previous equation we get,
\begin{eqnarray}\label{4.18}
\Big[4a^2(\Delta_r-a^2-\frac{r}{2}{\Delta_r}')+(2a^2-2\Delta_r+\frac{r}{2}{\Delta_r}')^2\Big]x^4+\Big[2(2a^2
-2\Delta_r+\frac{r}{2}{\Delta_r}')(\frac{r^3}{3}{\Delta_r}'\nonumber \\
 -r^2\Delta_r)-2a^2r^3{\Delta_r}'\Big]x^2+\big(\frac{r^3}{2}{\Delta_r}'-r^2\Delta_r\big)^2=0\quad
\end{eqnarray}
which is in the form of quadratic equation in $x^2$ i.e.
\begin{displaymath}
a(x^2)^2+b(x^2)+c \end{displaymath}

The discriminant $\frac{b^{2}-4\,a\,c}{4}$ of this equation is
\begin{displaymath}
8a^2r^4\Delta_r^2[a^2-\Delta_r+\frac{r}{2}{\Delta_r}']
\end{displaymath}
Thus we can find solution of equation (\ref{4.18}) as
\begin{equation}\label{4.19}
x^{2}=\frac{r^2\bigg(a\pm\sqrt{a^2-\Delta_r+\frac{r}{2}{\Delta_r}'}\bigg)^2}{\bigg(2\Delta_r-2a^2-\frac{r}{2}{\Delta_r}'\bigg)\mp2a\sqrt{a^2-\Delta_r+\frac{r}{2}{\Delta_r}'}}
\end{equation}
where,
\begin{displaymath}
\bigg(2\Delta_r-2a^2-\frac{r}{2}{\Delta_r}'\bigg)-4a^2\bigg(a^2-\Delta_r+\frac{r}{2}{\Delta_r}'\bigg)=Q_{+}\,Q_{-}
\end{displaymath}
and
\begin{equation}\label{4.20}
Q_{\pm}=\bigg(2\Delta_r-2a^2-\frac{r}{2}{\Delta_r}'\bigg)\pm2a\sqrt{a^2-\Delta_r+\frac{r}{2}{\Delta_r}'}
\end{equation}
Therefore,
\begin{equation}
x^{2}=\frac{r^2(\Delta_{r}-Q_{\mp})}{Q_{\mp}}
\end{equation}
From equations (\ref{4.19}) and (\ref{4.20}) the solution for $x$
thus takes the simple form
\begin{eqnarray}
x=-\frac{r}{\sqrt{Q_\mp}}\bigg(a\pm\sqrt{a^2-\Delta_r+\frac{r}{2}{\Delta_r}'}\bigg)
=\frac{r}{2a\sqrt{Q_\mp}}\big(\frac{r}{2}{\Delta_r}'-2\Delta_r+Q_\mp\big)
\end{eqnarray}

The upper sign in the equations apply to retrograde orbits and
lower sign apply to direct orbits.
\\
Using value of $x$ in equation (\ref{4.21}),
\begin{eqnarray}
E=\frac{1}{r\sqrt{Q_\mp}}\bigg(\Delta_r-a^2\mp
a\sqrt{a^2-\Delta_r+\frac{r}{2}{\Delta_r}'}\bigg)
=\frac{1}{2r\sqrt{Q_\mp}}\big(\frac{r}{2}{\Delta_r}'+Q_\mp\big)
\end{eqnarray}
and thus, $L=aE+x$ i,e;
\begin{equation}
L=\frac{1}{r\sqrt{Q_\mp}}\bigg[a(\Delta_r-a^2-r^2)\mp
(a^2+r^2)\sqrt{a^2-\Delta_r+\frac{r}{2}{\Delta_r}'}\,\bigg]
\end{equation}

Here $E$ and $L$ are the energy and the angular momentum per unit
mass, of a particle describing a circular orbit of reciprocal
radius $u$.\\
\\
Now, using the values of $L$, $x$ and $E$ to find velocity of test particle, we know \\
\begin{displaymath} \Omega=\frac{d\phi}{dt}=\frac
{\dot{\phi}}{\dot{t}}=\frac{\mp\sqrt{a^2-\Delta_r+\frac{r}{2}{\Delta_r}'}}{\Big(r^2\mp a\sqrt{a^2-\Delta_r+\frac{r}{2}{\Delta_r}'}\Big)}\end{displaymath}
and \begin{displaymath}
\omega=\frac{a}{\Sigma^2}\big[(r^2+a^2)\Delta_\theta-\Delta_r\big]\end{displaymath}
where $\Sigma^2=(r^2+a^2)^2\Delta_\theta-\Delta_r a^2
sin^2\theta$\\\\
The rotational velocity of test particle in the orbit around the
central mass is \begin{displaymath} V_\varphi =\frac{\Sigma^2
sin^2\theta}{\rho^2\sqrt{\Delta_r
\Delta_\theta}}(\Omega-\omega)\end{displaymath} Thus, using these
values rotational velocity we obtained as
\begin{equation}
V_{\varphi}=\mp\frac{\bigg[\big(r^2+a^2\big)\sqrt{a^2-\Delta_r+\frac{r}{2}{\Delta_r}'}\pm
a\big(r^2+a^2-\Delta_r\big)\bigg]}{\big(r^2\mp
a\sqrt{a^2-\Delta_r+\frac{r}{2}{\Delta_r}'}\big)\sqrt{\Delta_r}}
\end{equation}
In the limit $\Lambda \longrightarrow 0$ \begin{equation}
V_\varphi= \mp \frac{\sqrt{\frac{M}{r}}\Big[1+\frac{a^2}{r^2}\pm
\frac{2a}{r}\sqrt{\frac{M}{r}}\Big]}{\Big[1\mp
\frac{a}{r}\sqrt{\frac{M}{r}}\Big]\Big[1+\frac{a^2}{r^2}-\frac{2M}{r}\Big]^{1/2}}
\end{equation} which is for Kerr space-time metric.\\
Similarly, in the limit $a\longrightarrow 0$,
\begin{equation}\label{4.22}
V_{\varphi}=\sqrt{\frac{\frac{M}{r}-\frac{\Lambda
r^2}{3}}{1-\frac{2M}{r}-\frac{\Lambda r^2}{3}}}
\end{equation}
Which is the Schwarzschild de-Sitter limit.\\\\
 When $\frac{M}{r}<<1$, equation (\ref{4.22}) reduces to the usual Newtonian
limit, with $G$ = $c$ = 1,
\begin{displaymath}
V_\varphi=(\frac{M}{r})^{\frac{1}{2}}
\end{displaymath}
i.e;
\begin{displaymath}
V_\varphi=(\frac{G\,M}{c^{2}\,r})^{\frac{1}{2}}
\end{displaymath}

\section{Analysis of geodesic motion in Kerr-de Sitter space-time}
The time-like geodesics motion in Kerr-de Sitter space-time is one
of the complicated problem. The equation of motion in circular
orbit is a non linear. It has distance from center as independent
parameter and velocity as dependent parameter. It contains mass,
cosmological constant and angular momentum or spinning constant or
rotational Kerr parameter as constant parameters. For different values of
mass, cosmological constant and Kerr parameter; nature of geodesic motion can be displayed
which are as follows:
\\\\
1)The rotational velocity increases with mass (M) keeping cosmological constant
non-negative. For non-negative value of $\Lambda$; `$v$' vs `$r$'
curves do not meet to each other while they meet after a certain
point for negative value of $\Lambda$. The value of meeting point
depends upon the value of $\Lambda$. The plots of `$v$' vs `$r$'
curves are shown in fig. 2.
\begin{figure}
\vspace{0.0cm} \centering
\includegraphics[height=4.0cm,width=4.15cm]{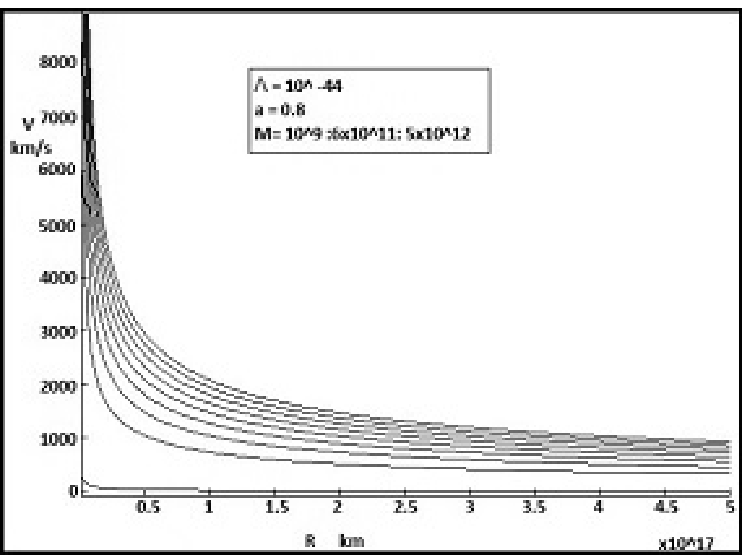}
\includegraphics[height=4.0cm,width=4.15cm]{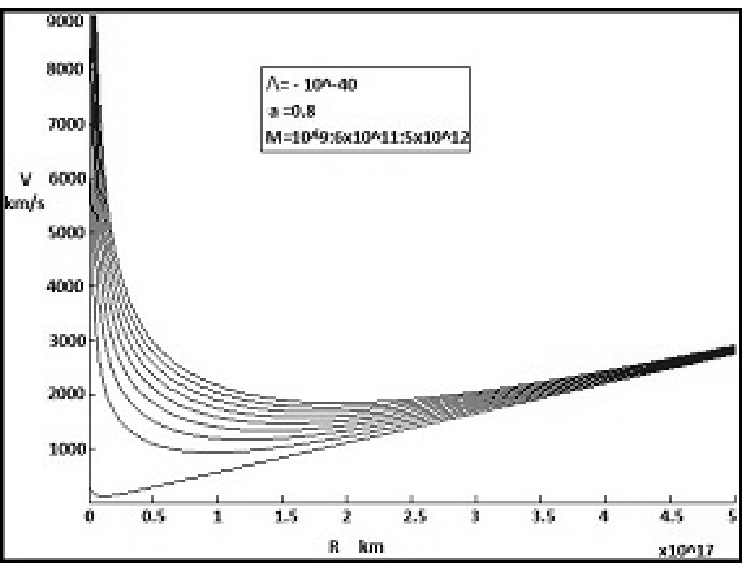}
\includegraphics[height=4.05cm,width=4.15cm]{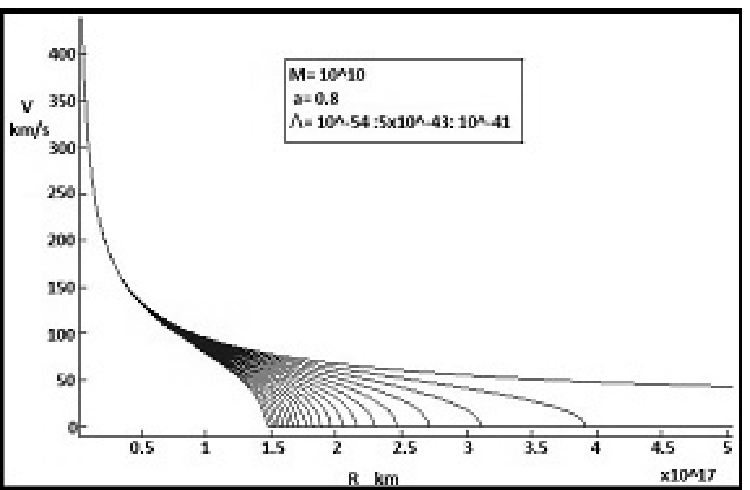}
\includegraphics[height=4.04cm,width=4.15cm]{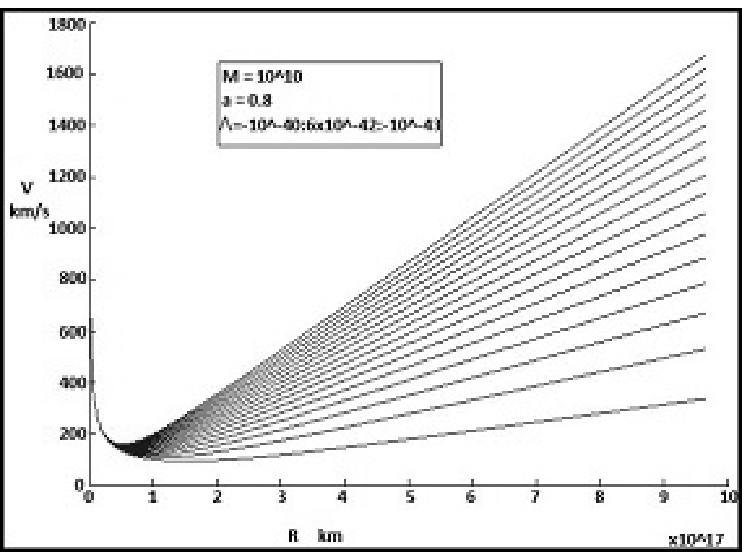}
\caption{ First two figures show geodesic curves for different values of mass keeping constant value of $a$ and constant positive and negative values of $\Lambda $ respectively while that of last two show geodesic curves of positive and negative variation of $\Lambda$ respectively with constant value of $M$ and $a$}
\end{figure}
\\\\
2)The curvature is dependent on value of $\Lambda$. For non negative
value of $\Lambda$ velocity decreases as distance increases while
velocity increases with distance for negative value of $\Lambda$
keeping mass constant. The plot showing different curves due to
the variation of $\Lambda$ is shown in fig. 2.
\\\\
3)Since $a$
appears as multiplicative factor with mass (M) and lies in the
range 0.1 $-$ 1.0, its effect in curvature is found to be
negligible as shown in fig. 3.
\begin{figure}
\vspace{0.0cm} \centering
\includegraphics[height=4.1cm,width=5.17cm]{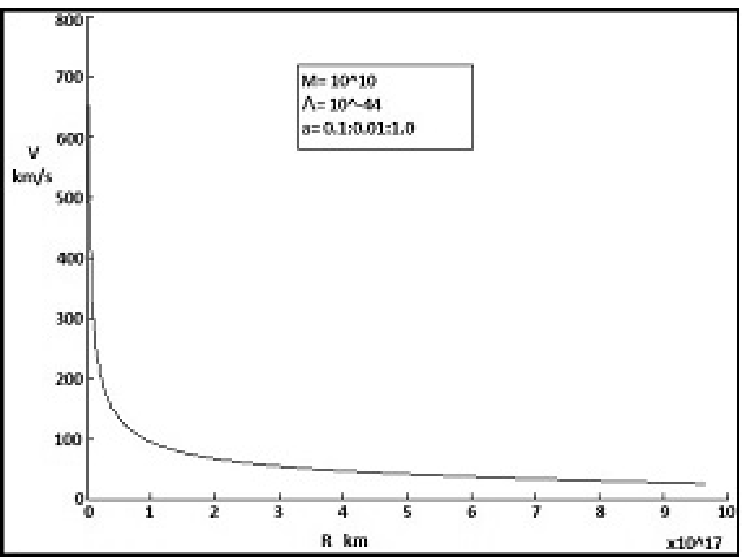}
\includegraphics[height=4.1cm,width=5.17cm]{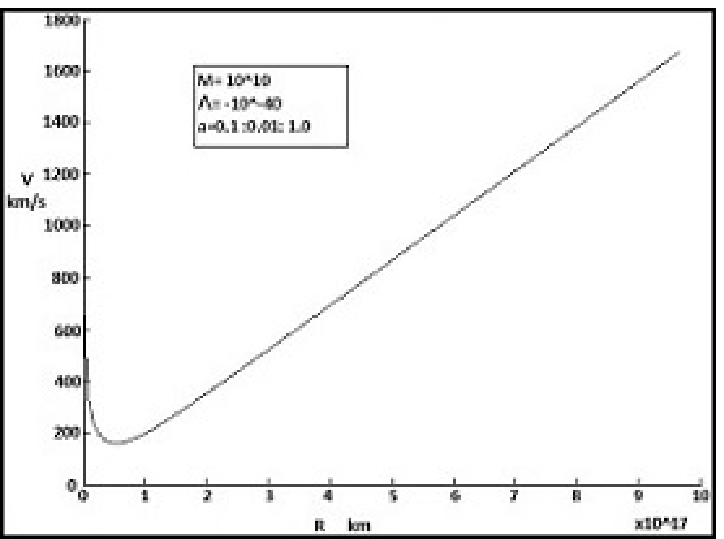}
\caption{Geodesic curves with variation of $a$ with constant positive and negative value of
$\Lambda $ and constant value of $M$.}
\end{figure}

\section{Application of geodesic motion in Kerr-de Sitter space-time}

In this section we intend whether geodesic motion in Kerr-de Sitter
space-time could be applied or not. So, we focused our
interest to the database of rotational curves data of galaxies which were well fitted to our
problems. For this we took non-linear curve fit statistics for
which R-square and adjusted R-square errors in such a way that it
should be non negative. We selected 30 galaxies and fitted the
values of mass (M), cosmological constant ($\Lambda$) and
rotational Kerr parameter ($a$). Among 30 galaxies, the values of
rotational velocity ($v$) and the distance from the galactic
center ($r$) for 28 galaxies are taken from the database provided
by Sofue et al. (2007) and that for the rest 2 galaxies (NGC 3379
and NGC 4100) are taken from the data digitized by Software
`Labfit' from the literatures of Brownstein \& Moffat (2005). We
used software Matlab7.6 to carry out non-liner least square curve
fit method to estimate the values of M, $\Lambda$ and $a$. In this
parametric curve fitting M, $\Lambda$ and $a$ were obtained as
unknown coefficients from our equation of motion.
\\
\\
Out of 30 galaxies we could fit positive cosmological constant for
23 galaxies while it was negative for 7 galaxies. To find the
unknown parameters we analyzed the nature of curve of equation of
motion in Kerr-de Sitter space-time and fitted with rotation
curves data of galaxies. As discussed in above section, it is found that the rotational velocity
in the flat portion in the rotation curves data of galaxies
increases with mass (M) keeping $\Lambda$ non negative. Similarly,
curvature is found to be dependent on the value of $\Lambda$. For
non negative value of $\Lambda$ velocity decreases as distance
increases while for negative value it increases as distance
increases keeping mass constant. We found value of $a$ has
negligible contribution in the curvature but it affects in
goodness of fit statistics. So we minimized the errors associated
with it. Thus we fitted rotation curves data for most appropriate
values of mass (M), cosmological constant ($\Lambda$)
and Kerr parameter ($a$).\\
\\
The value of cosmological constant is found to lie within the
range of $\rm (1.790\pm0.286)\times10^{-49}\,\, km^{-2}$ to $\rm
(7.523\pm1.204)\times10^{-42}\,\, km^{-2}$ for positive value of
$\Lambda$ while $\rm -(3.983\pm0.637)\times10^{-41} \,\,km^{-2}$
to $\rm -(1.860\pm0.298)\times10^{-42} \,\,km^{-2}$ for negative
value of $\Lambda$. Among positive values the least value is found
for the galaxy NGC 3521 and maximum is found for the galaxy NGC
3034. Similarly, among negative values maximum is found for NGC
2708 and minimum value value is found for NGC 3495. Negative
cosmological constants were found to fit for the galaxies which
are high red shifted spiral galaxies. Exception to these is NGC
4569 which is high blue shifted having radial velocity equals -235
km/s. This is found to be Sab morphological type and having LINER
activity. Galaxies having low value of radial velocities ($\sim$
-300 km/s to 1116 km/s) were found to fit with positive
cosmological constants. Exception to these are NGC 1097, NGC 1365,
NGC 4321 and NGC 4565 ($\sim$ 1230 km/s to 1636 km/s). We have
found greater value ($\rm \sim 10^{-42}\,\,km^{-2}$ to $\rm
10^{-47}\,\,km^{-2}$) of cosmological constants for galaxies NGC
1808, NGC 3034, NGC 3521, NGC 4736 and NGC 5194. Out of which NGC
1808, NGC 3034 are found to be of active galaxies and NGC 5194
(M51) shows peculiar characteristics (Sofue et al., 1999).
While NGC 4736 is found to be of Sab type and having radial
velocity equals 606 km/s. In general, we found cosmological
constant in the range of $\rm 10^{-41}\,\,km^{-2}$ to $\rm
10^{-49}\,\,km^{-2}$ for positive value of cosmological constant
and in the range of $-\rm 10^{-41}\,\,km^{-2}$ to $\rm
-10^{-42}\,\,km^{-2}$ for negative value of $\Lambda$ which are in
agreement with the values found in other literatures.\\
\\
The estimated mass of the galaxies lie in the range of $\rm
(0.12\pm0.02)\times10^{10}\,\, M_\odot$ to $\rm
(70.37\pm11.26)\times10^{10}\,\, M_\odot$ where $\rm 1M_\odot =
1.989\times10^{30}$ kg which are also in good agreement with other
estimated values found in literatures. Value of mass is found
small $\rm (0.73\pm0.12)\times 10^{10} \,\,M_\odot$ for NGC 3034
which is Ir II type and have positive value of $\Lambda$. While
value of mass $\rm (0.12\pm0.02)\times 10^{10} \,\,M_\odot$) is
found for NGC 3495 which is Sd type and have negative cosmological
constant. Greater value of mass is found for NGC 1097 which is of
SBb type. We had also calculated mass of elliptical galaxy NGC
3379 whose value is found to fit with $\rm
(1.027\pm0.164)\times10^{11}\,\,M_\odot$. In general mass is found
to lie in the range $\rm 10^{9}\,\,M_\odot$ to $\rm
10^{11}\,\,M_\odot$ for spiral barred and unbarred
galaxies.\\
\\
Similarly value of  Kerr parameter we fitted lie in the range of
(0.7044$\pm$0.1127) to (0.9990$\pm$0.1598). These are the values that could
have minimum error in non-linear least square curve fitting. Small value
(0.7044$\pm$0.1127) of $a$ is found for NGC 4736 which is early
type barred spiral (Sab) type and it has small mass while greater
value (0.9990$\pm$0.1598) is found for NGC 3379 which is of
elliptical type. Recent measurements of the Kerr parameters $a$
for two stellar sized black-hole binaries in our Galaxy (Shafee et
al. 2006) for GRO J1655-40 and 4U 1543-47 are estimated to fall in
the range $a=0.65-0.75$ and $a=0.75-0.85$, respectively. Our
estimated values are quite reasonable because we know that spin
angular momentum for a collective mass of the galaxy has always
higher value than for a black hole mass. We found most of our
estimated massive galaxies ($>50\%$) are fitted with the higher
value of $a$. Since spin of barred and unbarred galaxies not only
mass dependent but also depends on their local inner activities
such as starbursts as well as globular cause such as its
neighbouring galaxies, its position in galaxy cluster etc. These
results might be  interesting in the future studies.\\
\\
The best fitted graph of some galaxies are shown in fig.(5).
The detailed best fitted values of mass (M), cosmological constant
($\Lambda$), Kerr parameter ($a$) and errors associated with it
are given in Table (1).

\begin{table}[h]  \tiny
\centering \caption[]{Estimated values of Mass (M), Cosmological
Constant $\Lambda$ and Kerr Parameter ($a$) for different
Galaxies}
\begin{tabular} {|p{0.381in}|p{0.8in}|p{0.85in}|p{1.0in}|p{0.7in}|p{0.41in}|p{0.24in}|p{0.36in}|p{0.24in}|}
\hline Name Of  Galaxy & Mass  $\rm (M_\odot)$ & Mass In kg &
Cosmological Constant  $\Lambda $ (km$^{-2}$) & Kerr Parameter
($a$)  & SSE  & R- Square Error & Adjusted R-Squared Error & RMSE
\\ \hline NGC0224 & (39.05$\pm$6.25)$\times$10$^{10}$ &
(7.767$\pm$1.242)$\times$10$^{41}$ &
(2.339$\pm$0.374)$\times$10$^{-48}$
& (0.9900$\pm$0.1584) & 91319.635 & 0.3090 & 0.3090 & 13.190 \\
\hline NGC0891 & (18.88$\pm$3.02)$\times$10$^{10}$ &
(3.755$\pm$0.601)$\times$10$^{41}$ &
(2.337$\pm$0.374)$\times$10$^{-48}$ & (0.7800$\pm$0.1248) &
14462.815 & 0.9472 & 0.9472 & 6.021 \\ \hline NGC1097 & (70.37$\pm$11.26)$\times$10$^{10}$ &
(13.996$\pm$2.239)$\times$10$^{41}$ &
(2.331$\pm$0.373)$\times$10$^{-48}$ & (0.8900$\pm$0.1424) &
25440.000 & 0.0427 & 0.0427 & 10.520 \\ \hline NGC1365 &
(43.36$\pm$6.94)$\times$10$^{10}$ &
(8.624$\pm$1.380)$\times$10$^{41}$ &
(5.023$\pm$0.804)$\times$10$^{-49}$ & (0.7980$\pm$0.1277) &
17726.015 & 0.0165 & 0.0165 & 8.704 \\ \hline NGC1808 &
(9.55$\pm$1.53)$\times$10$^{10}$ &
(1.899$\pm$0.304)$\times$10$^{41}$ &
(2.020$\pm$0.323)$\times$10$^{-47}$ & (0.8809$\pm$0.1409) &
31935.047 & 0.7259 & 0.7359 & 11.990 \\ \hline  NGC2683 &
(11.69$\pm$1.87)$\times$10$^{10}$ &
(2.325$\pm$0.372)$\times$10$^{41}$ &
(2.330$\pm$0.373)$\times$10$^{-48}$ & (0.8600$\pm$0.1376) &
855.955 & 0.2592 & 0.2592 & 13.310
\\ \hline

NGC2903 & (25.04$\pm$4.01)$\times$10$^{10}$ &
(4.980$\pm$0.797)$\times$10$^{41}$ &
(2.278$\pm$0.364)$\times$10$^{-48}$ & (0.7721$\pm$0.1235) &
840.200 & 0.8869 & 0.8863 & 2.191 \\ \hline NGC3031 &
(16.06$\pm$2.56)$\times$10$^{10}$ &
(3.194$\pm$0.511)$\times$10$^{41}$ &
(2.337$\pm$0.374)$\times$10$^{-48}$ & (0.9500$\pm$0.1520) &
53039.950 & 0.7400 & 0.7400 & 12.850 \\ \hline NGC3034 &
(0.73$\pm$0.12)$\times$10$^{10}$ &
(0.145$\pm$0.023)$\times$10$^{41}$ &
(7.523$\pm$1.204)$\times$10$^{-42}$ & (0.9530$\pm$0.1525) &
2429.588 & 0.9843 & 0.9843 & 5.934 \\ \hline NGC3079 &
(21.76$\pm$3.48)$\times$10$^{10}$ &
(4.328$\pm$0.692)$\times$10$^{41}$ &
(4.100$\pm$0.656)$\times$10$^{-49}$ & (0.9600$\pm$0.1536) &
16568.786 & 0.8425 & 0.8425 & 7.107 \\ \hline NGC3379 &
(10.27$\pm$1.64)$\times$10$^{10}$ &
(2.043$\pm$0.327)$\times$10$^{41}$ &
(2.220$\pm$0.355)$\times$10$^{-48}$ & (0.9990$\pm$0.1598) &
12077.879 & 0.0494 & 0.0494 & 25.900 \\ \hline NGC3521 &
(21.76$\pm$3.48)$\times$10$^{10}$ &
(4.328$\pm$0.692)$\times$10$^{41}$ &
(1.790$\pm$0.286)$\times$10$^{-49}$ & (0.8720$\pm$0.1395) &
63070.305 & 0.0430 & 0.0430 & 15.090 \\ \hline NGC3628 &
(18.65$\pm$2.98)$\times$10$^{10}$ &
(3.709$\pm$0.593)$\times$10$^{41}$ &
(2.221$\pm$0.355)$\times$10$^{-48}$ & (0.8200$\pm$0.1312) &
14779.169 & 0.0639 & 0.0556 & 11.490 \\ \hline NGC4100 &
(13.96$\pm$2.23)$\times$10$^{10}$ &
(2.776$\pm$0.444)$\times$10$^{41}$ &
(3.130$\pm$0.501)$\times$10$^{-48}$ & (0.8800$\pm$0.1408) &
1390.000 & 0.0424 & 0.0424 & 10.760 \\ \hline NGC4321 &
(46.55$\pm$7.45)$\times$10$^{10}$ &
(9.259$\pm$1.481)$\times$10$^{41}$ &
(3.220$\pm$0.516)$\times$10$^{-48}$ & (0.7974$\pm$0.1276) &
29804.774 & 0.0282 & 0.2352 & 12.030 \\ \hline NGC4565 &
(49.47$\pm$7.92)$\times$10$^{10}$ &
(9.839$\pm$1.574)$\times$10$^{41}$ &
(2.342$\pm$0.375)$\times$10$^{-48}$ & (0.8200$\pm$0.1312) &
8539.000 & 0.9013 & 0.9013 & 2.729 \\ \hline NGC4631 &
(16.12$\pm$2.58)$\times$10$^{10}$ &
(3.206$\pm$0.513)$\times$10$^{41}$ &
(2.338$\pm$0.374)$\times$10$^{-48}$ & (0.7410$\pm$0.1186) &
2559.000 & 0.9633 & 0.9633 & 2.729 \\ \hline NGC4736 &
(6.89$\pm$1.10)$\times$10$^{10}$ &
(1.370$\pm$0.219)$\times$10$^{41}$ &
(1.242$\pm$0.198)$\times$10$^{-42}$ & (0.7044$\pm$0.1127) &
128.300 & 0.9969 & 0.9969 & 1.095 \\ \hline NGC5033 &
(61.73$\pm$9.88)$\times$10$^{10}$ &
(12.228$\pm$1.956)$\times$10$^{41}$ &
(7.640$\pm$1.222)$\times$10$^{-49}$ & (0.8600$\pm$0.1376) &
9858.784 & 0.7809 & 0.7809 & 4.952 \\ \hline NGC5194 &
(16.63$\pm$2.66)$\times$10$^{10}$ &
(3.308$\pm$0.529)$\times$10$^{41}$ &
(4.048$\pm$0.648)$\times$10$^{-42}$ & (0.7490$\pm$0.1198) &
2300.000 & 0.9934 & 0.9934 & 3.700 \\ \hline NGC5236 &
(24.42$\pm$3.91)$\times$10$^{10}$ &
(4.857$\pm$0.777)$\times$10$^{41}$ &
(2.296$\pm$0.367)$\times$10$^{-48}$ & (0.7679$\pm$0.1228) &
57164.049 & 0.0119 & 0.0097 & 11.090 \\ \hline NGC5457 &
(16.70$\pm$2.58)$\times$10$^{10}$ &
(3.321$\pm$0.531)$\times$10$^{41}$ &
(2.180$\pm$0.349)$\times$10$^{-49}$ & (0.7419$\pm$0.1187) &
784.106 & 0.9766 & 0.9766 & 2.447 \\ \hline NGC5907 &
(41.23$\pm$6.59)$\times$10$^{10}$ &
(8.200$\pm$1.312)$\times$10$^{41}$ &
(2.347$\pm$0.375)$\times$10$^{-48}$ & (0.8600$\pm$0.1376) &
3413.499 & 0.9643 & 0.9643 & 3.183 \\ \hline NGC2708 &
(30.76$\pm$4.92)$\times$10$^{10}$ &
(6.118$\pm$0.979)$\times$10$^{41}$ &
-(1.860$\pm$0.298)$\times$10$^{-42}$ & (0.7842$\pm$0.1255) &
3893.188 & 0.8909 & 0.8909 & 2.556 \\ \hline NGC3495 &
(0.12$\pm$0.02)$\times$10$^{10}$ &
(0.023$\pm$0.004)$\times$10$^{41}$ &
-(3.983$\pm$0.637)$\times$10$^{-41}$ & (0.9100$\pm$0.1456) &
15912.276 & 0.7329 & 0.7297 & 13.680 \\ \hline NGC3672 &
(4.08$\pm$0.65)$\times$10$^{10}$ &
(0.811$\pm$0.129)$\times$10$^{41}$ &
-(1.329$\pm$0.213)$\times$10$^{-41}$ & (0.8896$\pm$0.1423) &
4407.277 & 0.9130 & 0.9130 & 4.742 \\ \hline NGC4303 &
(4.57$\pm$0.73)$\times$10$^{10}$ &
(0.909$\pm$0.145)$\times$10$^{41}$ &
-(3.450$\pm$0.552)$\times$10$^{-42}$ & (0.9900$\pm$0.1584) &
383.910 & 0.3278 & 0.3225 & 1.739 \\ \hline NGC4569 &
(3.16$\pm$0.50)$\times$10$^{10}$ &
(0.629$\pm$0.101)$\times$10$^{41}$ &
-(3.970$\pm$0.635)$\times$10$^{-41}$ & (0.9860$\pm$0.1578) &
3900.010 & 0.9247 & 0.9239 & 6.547 \\ \hline NGC6951 &
(5.04$\pm$0.81)$\times$10$^{10}$ &
(1.002$\pm$0.160)$\times$10$^{41}$ &
-(1.783$\pm$0.285)$\times$10$^{-41}$ & (0.8550$\pm$0.137) &
3215.000 & 0.8449 & 0.8432 & 5.818 \\ \hline UGC3691 &
(0.40$\pm$0.06)$\times$10$^{10}$ &
(0.079$\pm$0.012)$\times$10$^{41}$ &
-(1.394$\pm$0.223)$\times$10$^{-41}$ & (0.9129$\pm$0.1461) &
6228.231 & 0.8795 & 0.8795 & 7.031 \\ \hline
\end{tabular}
\end{table}

\begin{figure}[h]
\vspace{0.0cm} \centering
\includegraphics[height=7.5cm,width=11.0cm]{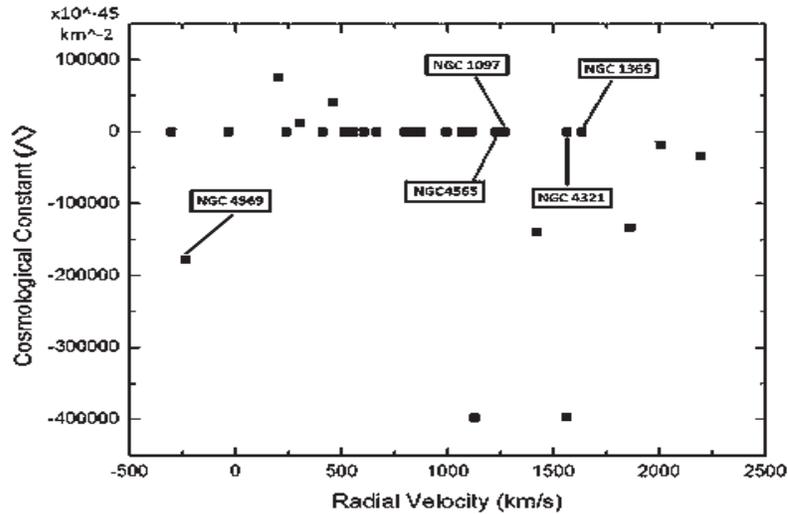}
\caption{Plot diagram of $\Lambda$ vs Radial Velocity of galaxies.
Each square box contains the name of galaxy having exceptional
behaviour. }
\end{figure}

\begin{figure}[h]
\vspace{0.0cm} \centering
\includegraphics[height=4.43cm,width=4.2cm]{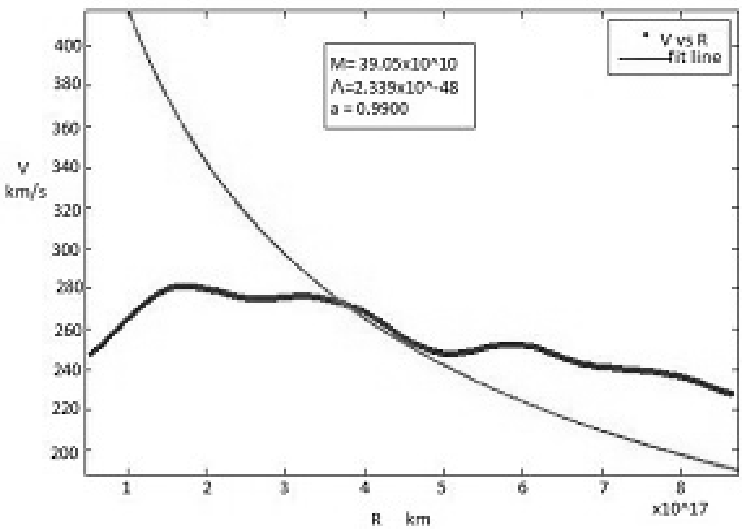}
\includegraphics[height=4.46cm,width=4.2cm]{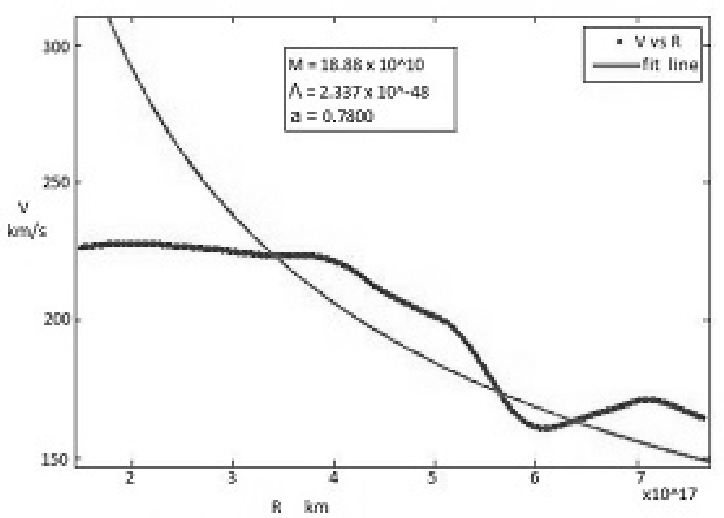}
\includegraphics[height=4.5cm,width=4.2cm]{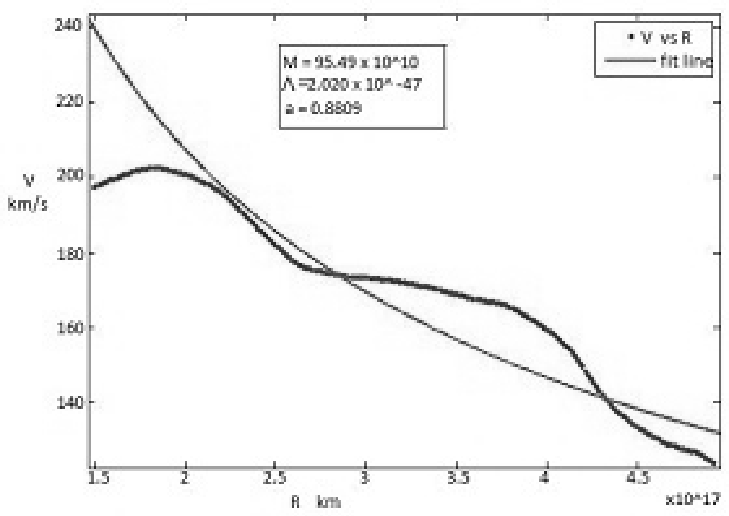}
\includegraphics[height=4.39cm,width=4.2cm]{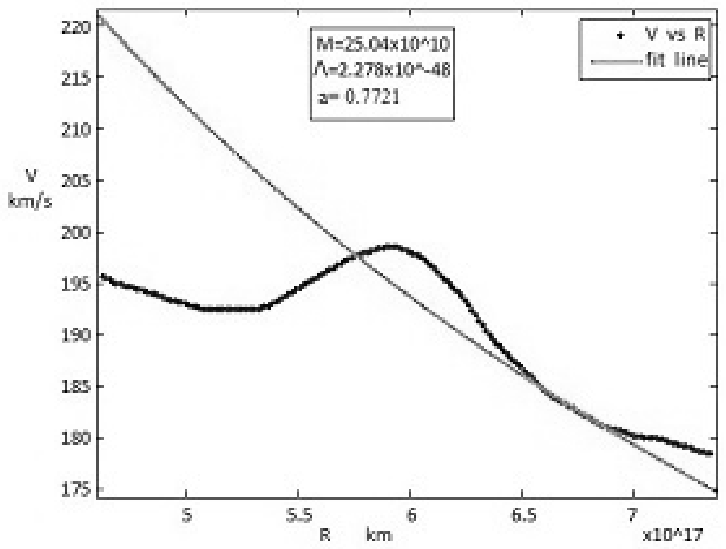}
\includegraphics[height=4.44cm,width=4.2cm]{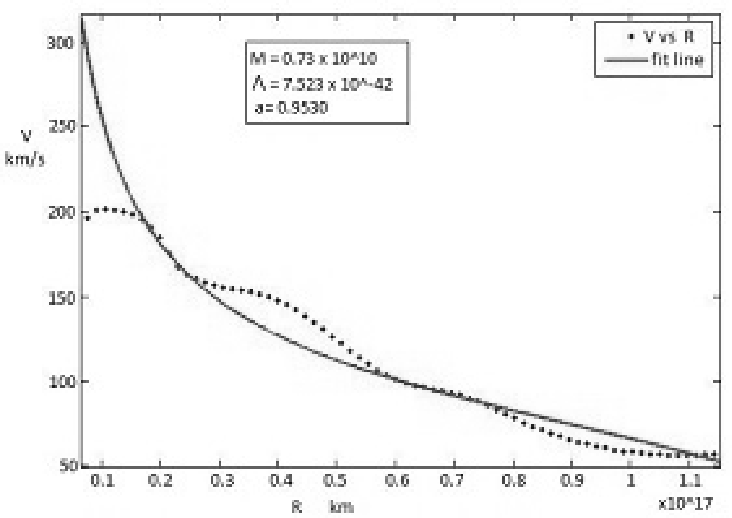}
\includegraphics[height=4.44cm,width=4.2cm]{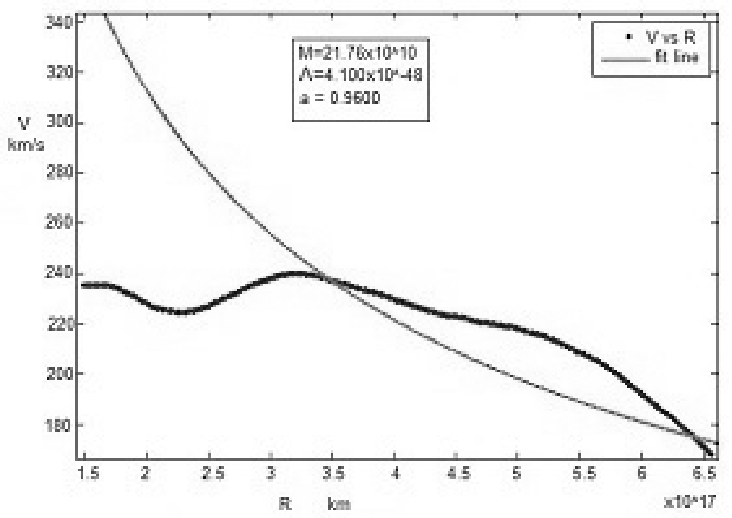}
\includegraphics[height=4.52cm,width=4.2cm]{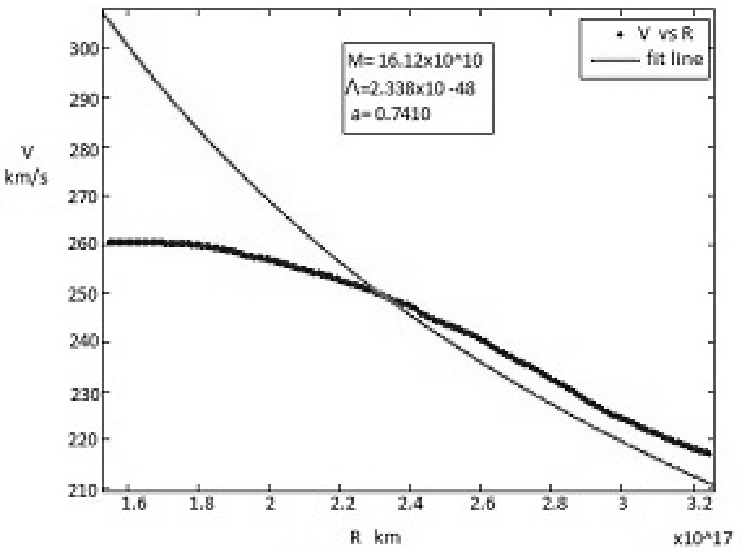}
\includegraphics[height=4.48cm,width=4.2cm]{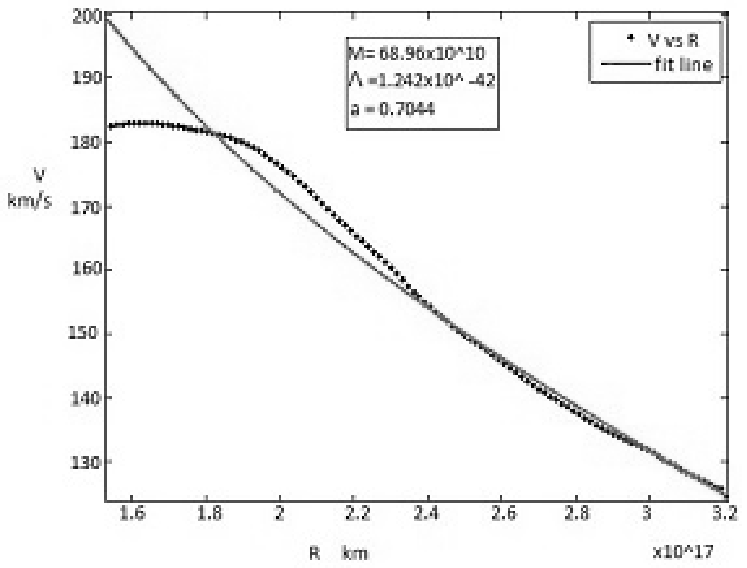}
\includegraphics[height=4.44cm,width=4.2cm]{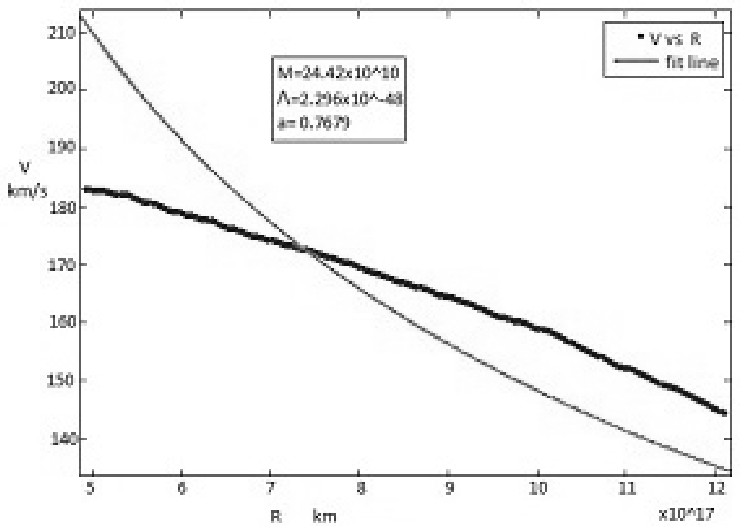}
\includegraphics[height=4.54cm,width=4.2cm]{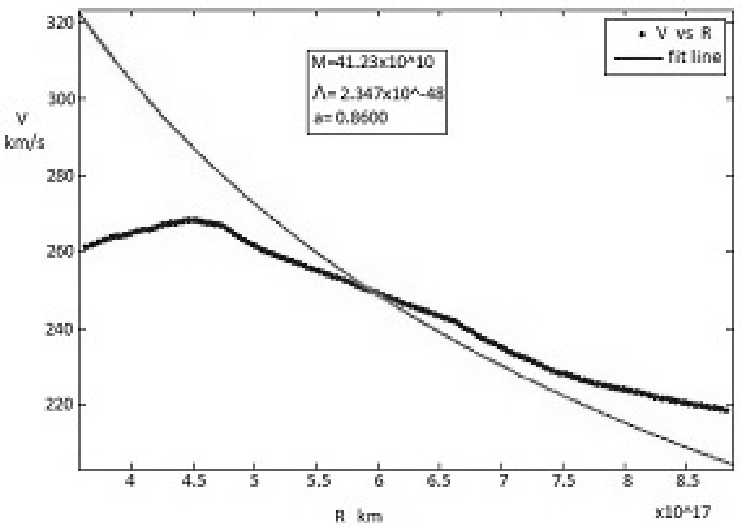}
\includegraphics[height=4.44cm,width=4.2cm]{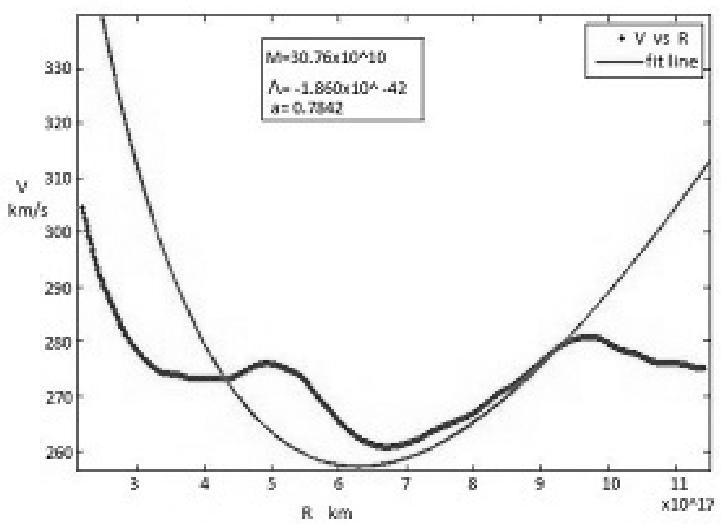}
\includegraphics[height=4.47cm,width=4.2cm]{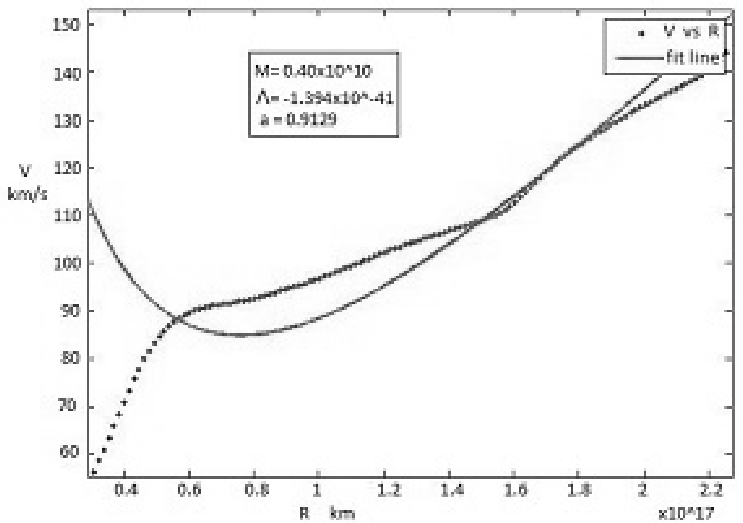}
\caption{The observed and fitted rotation curves of galaxies NGC
0224, NGC 0891, NGC 1808, NGC 2903, NGC
3034, NGC 3079, NGC 4631, NGC 4736, NGC
5236, NGC 5907, NGC 2708 and UGC 3691.. The thick solid and thin solid curve represents
the observed and fitted line respectively. The value of M,
$\Lambda$ and $a$ is shown inside the square box. }
\end{figure}
\section{Conclusion}

For some typical values of the parameters $\Lambda$, $D$, $L$ and $a$ there exists a potential-energy curve.
The minima in the potentials corresponds to the stable circular orbits while
maxima corresponds to unstable circular orbits. At the point of inflection the
last stable circular orbit occurs.\\
\\
The value of the rotational velocity in the flat portion in
the rotation curves data of galaxies is found to be increased with
mass (M) keeping cosmological constant ($\Lambda$) non-negative.
But, that for a constant negative value of $\Lambda$, `$v$' vs `$r$' curves
meet after a certain distance.
The meeting point is observed to be dependent on the value of
$\Lambda$.\\
\\
The Curvature of rotation curve data of Galaxies is found to
be dependent on the value of $\Lambda$ keeping mass constant. For
non negative value of $\Lambda$, velocity ($v$) decreases as
distance from galactic center ($r$) increases while for negative
value of $\Lambda$ velocity ($v$) increases as the distance increases.
Thus $\Lambda$ is found to be a essential parameter to fit the curvature of
rotation curves data of galaxies.\\
\\
Since Kerr parameter lies in the range of 0.1 to 1.0 and
appears as coefficient of mass in our equation of motion, it has
negligible contribution in `$v$' vs `$r$' curve. But it has an
effect in goodness of fit statistics, particularly in non-linear
least square curve fitting.
\\\\
The value of mass (M) of galaxies is estimated in the range of
$\rm (0.12\pm0.02)\times10^{10}\,\,M_\odot$ to $\rm
(70.37\pm11.26)\times10^{10}\,\,M_\odot$. NGC 1097 is found to be
more massive than others which is of SBb type. Least mass $\rm
(0.12\pm0.02)\times10^{10}M_\odot$ is found for NGC 3495 which is
fitted for negative cosmological constant and is of Sd type. Value
of mass is found small $\rm (0.73\pm0.12)\times10^{10}M_\odot$ for
NGC 3495 which is Ir II type. In general mass is found to lie in
the range $\rm 10^{9}\,\, M_\odot$ to $\rm 10^{11}\,\, M_\odot$
for spiral barred and unbarred galaxies.
\\\\
The nature of cosmological constant fitted for galaxies were
found to depend upon radial velocities of galaxies. Discarding
exceptional cases, for the higher values of radial velocity
cosmological constants were found to be negative while positive
for small values of RV. This suggest some local phase transition
effects at the time when the high redshift galaxies were formed.
\\\\
The value of Cosmological Constant is found to fall within
the range of $\rm (1.790\pm0.286)\times10^{-49}\,\, km^{-2}$ to
$\rm (7.523\pm1.204)\times10^{-42}\,\, km^{-2}$ for positive value
of $\Lambda$ while $\rm -(3.983\pm0.637)\times10^{-41}
\,\,km^{-2}$ to $\rm -(1.860\pm0.298)\times10^{-42}\,\, km^{-2}$
for negative value of $\Lambda$. Most of the galaxies were fitted
for the values of $\Lambda$ in the range of $\rm 10^{-49}
\,\,km^{-2}$ to $\rm 10^{-48}\,\, km^{-2}$.
\\\\
The value of Kerr parameter lies in the range of
(0.7044$\pm$0.1127) to (0.9990$\pm$0.1598). Small value, i.e.,
(0.7044$\pm$0.1127) of $a$ is found for NGC 4736 which is  Sab
type galaxy. It has small mass while greater value
(0.9990$\pm$0.1598) is found for NGC 3379 which is elliptical
type.

\section{Acknowledgement}

We are indebted to all faculty members and students at Central Department of Physics, Tribhuvan
University, Kirtipur for their constant helps and suggestions during the work. Specially, we would like to
thank Mr. P. R. Dhungel for his kind support.

\end{document}